\documentclass[lettersize,journal]{IEEEtran}
\usepackage{url}
\usepackage[utf8]{inputenc}
\usepackage{xcolor}
\usepackage{amsmath}
\DeclareMathOperator*{\argmax}{arg\,max}

\usepackage{multirow}
\usepackage{amssymb}
\usepackage{cite}
\usepackage{enumitem}
\usepackage{bm}
\usepackage{graphicx}
\usepackage{amsthm}
\newtheorem{thm}{Theorem}
\newtheorem{lem}[thm]{Lemma}

\usepackage{gensymb}

\usepackage[acronyms,nonumberlist,nopostdot,nomain,nogroupskip]{glossaries}
\usepackage{tablefootnote}
\usepackage{booktabs}
\usepackage{tabularx}
\usepackage{epsfig}
\usepackage[outdir=images_final/]{epstopdf}
\usepackage{tikz}
\usepackage{pgfplots}
\pgfplotsset{compat=newest} 
\pgfplotsset{plot coordinates/math parser=false}

\IEEEoverridecommandlockouts
\newcommand\copyrightnotice{%
	\begin{tikzpicture}[remember picture,overlay]
		\node[anchor=south,yshift=7.5pt] at (current page.south) {\fbox{\parbox{\dimexpr\textwidth-\fboxsep-\fboxrule\relax}{
					\footnotesize \textcopyright 2023 IEEE. Personal use of this material is permitted.
					Permission from IEEE must be obtained for all other uses, in any current or future media,
					including reprinting/republishing this material for advertising or promotional purposes,
					creating new collective works, for resale or redistribution to servers or lists,
					or reuse of any copyrighted component of this work in other works.}}};
	\end{tikzpicture}
}

\usetikzlibrary{plotmarks,patterns,patterns.meta,decorations.pathreplacing,backgrounds,calc,arrows,arrows.meta,spy,matrix}
\usepgfplotslibrary{patchplots,groupplots}
\usepackage{tikzscale}
\usepackage{siunitx}

\usepackage{multirow}
\usepackage{algorithm2e, setspace}
\SetAlCapNameFnt{\footnotesize}
\SetAlCapFnt{\footnotesize}
\RestyleAlgo{ruled}
\SetKwComment{Comment}{/* }{ */}
\usepackage{algpseudocode}

\usepackage[font=scriptsize]{subcaption}
\usepackage[font=footnotesize]{caption}

\usepackage{mathtools}

\usepackage{dblfloatfix}    % To enable figures at the bottom of page
\usepackage{colortbl}

\newacronym{3gpp}{3GPP}{3rd Generation Partnership Project}
\newacronym{adc}{ADC}{Analog to Digital Converter}
\newacronym{5g}{5G}{5th generation}
\newacronym{6g}{6G}{6th generation}
\newacronym{aimd}{AIMD}{Additive Increase Multiplicative Decrease}
\newacronym{am}{AM}{Acknowledged Mode}
\newacronym{amc}{AMC}{Adaptive Modulation and Coding}
\newacronym{aqm}{AQM}{Active Queue Management}
\newacronym{awgn}{AGWN}{Additive White Gaussian Noise}
\newacronym{balia}{BALIA}{Balanced Link Adaptation}
\newacronym{bdp}{BDP}{Bandwidth-Delay Product}
\newacronym{bf}{BF}{beamforming}
\newacronym{cc}{CC}{Congestion Control}
\newacronym{cdf}{CDF}{Cumulative Distribution Function}
\newacronym{cn}{CN}{Core Network}
\newacronym{cqi}{CQI}{Channel Quality Information}
\newacronym{cp}{CP}{Control Plane}
\newacronym{csirs}{CSI-RS}{Channel State Information - Reference Signal}
\newacronym{dc}{DC}{Dual Connectivity}
\newacronym{rb}{RB}{Resource Block}
\newacronym{dce}{DCE}{Direct Code Execution}
\newacronym{dci}{DCI}{Downlink Control Information}
\newacronym{udp}{UDP}{User Datagram Protocol}
\newacronym{dl}{DL}{downlink}
\newacronym{fcfs}{FCFS}{first-come-first-served}
\newacronym{dmr}{DMR}{Deadline Miss Ratio}
\newacronym{fspl}{FSPL}{free-space path loss}
\newacronym{dmrs}{DMRS}{DeModulation Reference Signal}
\newacronym{e2e}{E2E}{End-to-End}
\newacronym{ppp}{PPP}{Poission Point Process}
\newacronym{aoi}{AoI}{Area of Interest}
\newacronym{cpu}{CPU}{Central Processing Unit}
 \newacronym{gpu}{GPU}{Graphics Processing Unit}
 \newacronym{tpu}{TPU}{Tensor Processing Unit}
\newacronym{si}{SI}{Study Item}
\newacronym{ecn}{ECN}{Explicit Congestion Notification}
\newacronym{edf}{EDF}{Earliest Deadline First}
\newacronym{enb}{eNB}{eNodeB}
\newacronym{epc}{EPC}{Evolved Packet Core}
\newacronym{es}{ES}{Edge Server}
\newacronym{cav}{CAV}{Connected and Autonomous Vehicle}
\newacronym{fdma}{FDMA}{Frequency Division Multiple Access}
\newacronym{fdd}{FDD}{Frequency Division Duplexing}
\newacronym{upa}{UPA}{Uniform Planar Array}
\newacronym[firstplural=Radio Access Technologies (RATs)]{rat}{RAT}{Radio Access Technology}
\newacronym[firstplural=Radio Access Technology (RTs)]{rt}{RT}{Radio Technology}
\newacronym{fs}{FS}{Fast Switching}
\newacronym{isd}{ISD}{inter-site distance}
\newacronym{ftp}{FTP}{File Transfer Protocol}
\newacronym{gnb}{gNB}{Next Generation Node Base}
\newacronym{harq}{HARQ}{Hybrid Automatic Repeat reQuest}
\newacronym{hetnet}{HetNet}{Heterogeneous Network}
\newacronym{hh}{HH}{Hard Handover}
\newacronym{hol}{HOL}{Head-of-Line}
\newacronym{ia}{IA}{Initial Access}
\newacronym{imt}{IMT}{International Mobile Telecommunication}
\newacronym{iot}{IoT}{Internet of Things}
\newacronym{los}{LOS}{Line of Sight}
\newacronym{lte}{LTE}{Long Term Evolution}
\newacronym{m2m}{M2M}{Machine to Machine}
\newacronym{mac}{MAC}{Medium Access Control}
\newacronym{mc}{MC}{Multi-Connectivity}
\newacronym{mcs}{MCS}{Modulation and Coding Scheme}
\newacronym{mec}{MEC}{Mobile Edge Cloud}
\newacronym{mi}{MI}{Mutual Information}
\newacronym{mimo}{MIMO}{Multiple Input Multiple Output}
\newacronym{mmwave}{mmWave}{millimeter wave}
\newacronym{mptcp}{MPTCP}{Multipath TCP}
\newacronym{mr}{MR}{Maximum Rate}
\newacronym{mss}{MSS}{Maximum Segment Size}
\newacronym{mtd}{MTD}{Machine-Type Device}
\newacronym{mtu}{MTU}{Maximum Transmission Unit}
\newacronym{nfv}{NFV}{Network Function Virtualization}
\newacronym{vnf}{VNF}{Virtualization Network Function}
\newacronym{gv}{GV}{ground vehicle}
\newacronym{vec}{VEC}{Vehicular Edge Computing}
\newacronym{sdn}{SDN}{Software Defined Networking}
\newacronym{nlos}{NLOS}{Non Line of Sight}
\newacronym{nlosb}{NLOSb}{Building Non Line of Sight}
\newacronym{nlosv}{NLOSv}{Vehicle Non Line of Sight}
\newacronym{nr}{NR}{New Radio}
\newacronym{ofdm}{OFDM}{Orthogonal Frequency Division Multiplexing}
\newacronym{pdcch}{PDCCH}{Physical Downlonk Control Channel}
\newacronym{pdcp}{PDCP}{Packet Data Convergence Protocol}
\newacronym{pdsch}{PDSCH}{Physical Downlink Shared Channel}
\newacronym{pdu}{PDU}{Packet Data Unit}
\newacronym{pf}{PF}{Proportional Fair}
\newacronym{pgw}{PGW}{Packet Gateway}
\newacronym{phy}{PHY}{Physical}
\newacronym{pbch}{PBCH}{Physical Broadcast Channel}
\newacronym[plural=\gls{mme}s,firstplural=Mobility Management Entities (MMEs)]{mme}{MME}{Mobility Management Entity}
\newacronym{prb}{PRB}{Physical Resource Block}
\newacronym{pss}{PSS}{Primary Synchronization Signal}
\newacronym{pucch}{PUCCH}{Physical Uplink Control Channel}
\newacronym{pusch}{PUSCH}{Physical Uplink Shared Channel}
\newacronym{rach}{RACH}{Random Access Channel}
\newacronym{ran}{RAN}{Radio Access Network}
\newacronym{red}{RED}{Random Early Detection}
\newacronym{rf}{RF}{Radio Frequency}
\newacronym{rlc}{RLC}{Radio Link Control}
\newacronym{rlf}{RLF}{Radio Link Failure}
\newacronym{rrc}{RRC}{Radio Resource Control}
\newacronym{rrm}{RRM}{Radio Resource Management}
\newacronym{rr}{RR}{Round Robin}
\newacronym{rs}{RS}{Remote Server}
\newacronym{rsrp}{RSRP}{Reference Signal Received Power}
\newacronym{rss}{RSS}{Received Signal Strength}
\newacronym{rtt}{RTT}{Round Trip Time}
\newacronym{rw}{RW}{Receive Window}
\newacronym{rx}{RX}{Receiver}
\newacronym{sa}{SA}{standalone}
\newacronym{sack}{SACK}{Selective Acknowledgment}
\newacronym{sap}{SAP}{Service Access Point}
\newacronym{sch}{SCH}{Secondary Cell Handover}
\newacronym{scoot}{SCOOT}{Split Cycle Offset Optimization Technique}
\newacronym{sdma}{SDMA}{Spatial Division Multiple Access}
\newacronym{sinr}{SINR}{Signal to Interference plus Noise Ratio}
\newacronym{sm}{SM}{Saturation Mode}
\newacronym{snr}{SNR}{Signal to Noise Ratio}
\newacronym{son}{SON}{Self-Organizing Network}
\newacronym{ss}{SS}{Synchronization Signal}
\newacronym{srs}{SRS}{Sounding Reference Signal}
\newacronym{sss}{SSS}{Secondary Synchronization Signal}
\newacronym{tb}{TB}{Transport Block}
\newacronym{tcp}{TCP}{Transmission Control Protocol}
\newacronym{tdd}{TDD}{Time Division Duplexing}
\newacronym{tdma}{TDMA}{Time Division Multiple Access}
\newacronym{tfl}{TfL}{Transport for London}
\newacronym{tm}{TM}{Transparent Mode}
\newacronym{prr}{PRR}{Packet Reception Ratio}
\newacronym{trp}{TRP}{Transmitter Receiver Pair}
\newacronym{tti}{TTI}{Transmission Time Interval}
\newacronym{ttt}{TTT}{Time-to-Trigger}
\newacronym{tx}{TX}{Transmitter}
\newacronym{ue}{UE}{User Equipment}
\newacronym{ul}{UL}{uplink}
\newacronym{uml}{UML}{Unified Modeling Language}
\newacronym{um}{UM}{Unacknowledged Mode}
\newacronym{utc}{UTC}{Urban Traffic Control}
\newacronym{vm}{VM}{Virtual Machine}
\newacronym{rsrq}{RSRQ}{Reference Signal Received Quality}
\newacronym{rssi}{RSSI}{Received Signal Strength Indicator}
\newacronym{crs}{CRS}{Cell Reference Signal}
\newacronym{v2v}{V2V}{Vehicle-to-Vehicle}
\newacronym{v2i}{V2I}{Vehicle-to-Infrastructure}
\newacronym{v2n}{V2N}{Vehicle-to-Network}
\newacronym{v2x}{V2X}{Vehicle-to-Everything}
\newacronym{vn}{VN}{Vehicular Node}
\newacronym{dsrc}{DSRC}{Dedicated Short Range Communication}
\newacronym{ci}{CI}{context information}
\newacronym{voi}{VoI}{value of information}
\newacronym{gps}{GPS}{Global Positioning System}
\newacronym{qos}{QoS}{Quality of Service}
\newacronym{qoe}{QoE}{Quality of Experience}
\newacronym{ml}{ML}{Machine Learning}
\newacronym{ahp}{AHP}{Analytic Hierarchy Process}
\newacronym{lidar}{LIDAR}{Light Detection and Ranging}
\newacronym{sumo}{SUMO}{Simulation of Urban MObility}
\newacronym{wave}{WAVE}{Wireless Access in Vehicular Environment}
\newacronym{c-its}{C-ITS}{Connected Intelligent Transportation System}
\newacronym{dash}{DASH}{Dynamic Adaptive Streaming over HTTP}
\newacronym{http}{HTTP}{HyperText Transfer Protocol}
\newacronym{nt}{NT}{Non-Terrestrial}
\newacronym{ntc}{NTC}{non-terrestrial communication}
\newacronym{ntn}{NTN}{Non-Terrestrial Network}
\newacronym{haps}{HAPS}{High Altitude Platform Station}
\newacronym{hap}{HAP}{High Altitude Platform}
\newacronym{leo}{LEO}{Low Earth Orbit}
\newacronym{meo}{MEO}{Medium Earth Orbit}
\newacronym{geo}{GEO}{Geostationary Earth Orbit}
\newacronym{uav}{UAV}{Unmanned Aerial Vehicle}
\newacronym{nsat}{nSAT}{Nanosatellite}
\newacronym{ehf}{EHF}{extremely high-frequency}
\newacronym{ioe}{IoE}{Internet of Everyone}
\newacronym{gan}{GaN}{Gallium Nitride}

\usepackage{tikz}
\usepackage{pgfplots}
\usepackage{tikzscale}
\usepackage{tikz-qtree}

\pgfplotsset{compat=newest}
\pgfplotsset{plot coordinates/math parser=false}
\pgfplotsset{every axis/.append style={
                    label style={font=\scriptsize},
                    tick label style={font=\scriptsize},
                    legend style={font=\scriptsize}
                    }}
\usetikzlibrary{plotmarks,shapes,patterns,decorations.pathreplacing,backgrounds,calc,arrows,arrows.meta,spy,matrix,shadows,trees,positioning,fit}
\usepgfplotslibrary{patchplots,groupplots}

\tikzstyle{startstop} = [rectangle, rounded corners, minimum width=2cm, minimum height=0.5cm,text centered, draw=black]
\tikzstyle{io} = [trapezium, trapezium left angle=70, trapezium right angle=110, minimum width=3cm, minimum height=1cm, text centered, draw=black]
\tikzstyle{process} = [rectangle, minimum width=2cm, minimum height=0.5cm, text centered, draw=black, alignb=center]
\tikzstyle{decision} = [ellipse, minimum width=2cm, minimum height=1cm, text centered, draw=black]
\tikzstyle{arrow} = [thick,<->,>=stealth]
\tikzstyle{line} = [thick,>=stealth]
\tikzstyle{darrow} = [thick,<->,>=stealth,dashed]
\tikzstyle{sarrow} = [thick,->,>=stealth]
\tikzstyle{larrow} = [line width=0.1mm,dashdotted,->,>=stealth]

\pgfkeys{/pgf/number format/.cd,1000 sep={\,}} % thousand separator: space

\makeatletter
\def\grd@save@target#1{%
  \def\grd@target{#1}}
\def\grd@save@start#1{%
  \def\grd@start{#1}}
\tikzset{
  grid with coordinates/.style={
    to path={%
      \pgfextra{%
        \edef\grd@@target{(\tikztotarget)}%
        \tikz@scan@one@point\grd@save@target\grd@@target\relax
        \edef\grd@@start{(\tikztostart)}%
        \tikz@scan@one@point\grd@save@start\grd@@start\relax
        \draw[minor help lines] (\tikztostart) grid (\tikztotarget);
        \draw[major help lines] (\tikztostart) grid (\tikztotarget);
        \grd@start
        \pgfmathsetmacro{\grd@xa}{\the\pgf@x/1cm}
        \pgfmathsetmacro{\grd@ya}{\the\pgf@y/1cm}
        \grd@target
        \pgfmathsetmacro{\grd@xb}{\the\pgf@x/1cm}
        \pgfmathsetmacro{\grd@yb}{\the\pgf@y/1cm}
        \pgfmathsetmacro{\grd@xc}{\grd@xa + \pgfkeysvalueof{/tikz/grid with coordinates/major step x}}
        \pgfmathsetmacro{\grd@yc}{\grd@ya + \pgfkeysvalueof{/tikz/grid with coordinates/major step y}}
        \foreach \x in {\grd@xa,\grd@xc,...,\grd@xb}
        \node[anchor=north] at (\x,\grd@ya) {\pgfmathprintnumber{\x}};
        \foreach \y in {\grd@ya,\grd@yc,...,\grd@yb}
        \node[anchor=east] at (\grd@xa,\y) {\pgfmathprintnumber{\y}};
      }
    }
  },
  minor help lines/.style={
    help lines,
    gray,
    line cap =round,
    xstep=\pgfkeysvalueof{/tikz/grid with coordinates/minor step x},
    ystep=\pgfkeysvalueof{/tikz/grid with coordinates/minor step y}
  },
  major help lines/.style={
    help lines,
    line cap =round,
    line width=\pgfkeysvalueof{/tikz/grid with coordinates/major line width},
    xstep=\pgfkeysvalueof{/tikz/grid with coordinates/major step x},
    ystep=\pgfkeysvalueof{/tikz/grid with coordinates/major step y}
  },
  grid with coordinates/.cd,
  minor step x/.initial=.5,
  minor step y/.initial=.2,
  major step x/.initial=1,
  major step y/.initial=1,
  major line width/.initial=1pt,
}
\makeatother

\newlength\fheight
\newlength\fwidth

\definecolor{steelblue}{RGB}{176,196,222}

\usepackage{hyperref}
\usepackage[capitalize]{cleveref}
\crefname{section}{Sec.}{Secs.}
\usetikzlibrary{decorations}

\makeglossaries
\glsdisablehyper
\begin{document}
\bstctlcite{IEEEexample:BSTcontrol}

\title{Real-Time HAP-Assisted Vehicular Edge Computing for Rural Areas}

\author{{{Alessandro Traspadini},~\IEEEmembership{Student Member, IEEE},
        {Marco Giordani},~\IEEEmembership{Member, IEEE},\\
        {Giovanni Giambene},~\IEEEmembership{Senior Member, IEEE},
        {Michele Zorzi},~\IEEEmembership{Fellow, IEEE}}
        \thanks{Alessandro Traspadini, Marco Giordani and Michele Zorzi are with the Department of Information Engineering, University of Padova, Padova, Italy (emails: \texttt{\{traspadini, giordani, zorzi\}@dei.unipd.it}). \newline
        Giovanni Giambene is with the Department of Information Engineering and Mathematical Sciences, University of Siena, Siena, Italy (email: \texttt{giovanni.giambene@unisi.it}).}
        \vspace{-0.6cm}
}

\maketitle
\copyrightnotice

\begin{abstract}
	\Glspl{ntn} are expected to be a key component of \gls{6g} networks to support broadband seamless Internet connectivity and expand the coverage even in rural and remote areas. In this context, \glspl{hap} can act as edge servers to process computational tasks offloaded by energy-constrained terrestrial devices such as \gls{iot} sensors and \glspl{gv}. In this paper, we analyze the opportunity to support \gls{vec} via \gls{hap} in a rural scenario where \glspl{gv} can decide whether to process data onboard or offload them to a \gls{hap}. We characterize the system as a set of queues in which computational tasks arrive according to a Poisson arrival process. 
	Then, we assess the optimal VEC offloading factor to maximize the probability of real-time service, given latency and computational capacity~constraints.
	
\end{abstract}

% arXiv
\begin{tikzpicture}[remember picture,overlay]
	\node[anchor=north,yshift=-10pt] at (current page.north) {\parbox{\dimexpr\textwidth-\fboxsep-\fboxrule\relax}{
			\centering\footnotesize This paper has been accepted for publication at IEEE Wireless Communications Letters (WCL). \textcopyright 2023 IEEE.\\
			Please cite it as: A. Traspadini, M. Giordani, G. Giambene and M. Zorzi, "Real-Time HAP-Assisted Vehicular Edge Computing for Rural Areas," in IEEE Wireless Communications Letters, doi: 10.1109/LWC.2023.3238851.}};
\end{tikzpicture}

\glsresetall

\section{Introduction}
\label{sec:intro}
One of the grand objectives of \gls{6g} wireless networks~\cite{giordani2020towards} is to enhance broadband Internet coverage in remote areas~\cite{Chaoub20216g}. 
The main issues towards this goal are the high costs of terrestrial deployments (especially for fiber-optic backhaul roll-out) as well as the lack of power sources and transport connectivity in rural regions. 
To overcome these problems, 6G will promote \glspl{ntn} with \glspl{uav}, \glspl{hap}, and satellites to provide fast connectivity thanks to their flexibility and inherent support for large coverage~\cite{giordani2021non}. %\cite{loonlibrary}.
For example, in 2020 a network of \glspl{hap} was deployed by Loon to bring Internet connectivity to unserved regions in Kenya. Similarly, in 2021, Airbus has proved the suitability of the  solar-powered Zephyr HAP to provide direct-to-device connectivity in the rural areas of Arizona, US.

 %provide broadband internet to the world by combining the persistence of satellites and HAPs with the flexibility of UAVs, such as Airbus Zephyr’s initiative. 

%Some practical projects have been developed and have already obtained promising results.
%Loon's project, for instance, consists of a network of \glspl{hap}  that fly at around 20~km and can provide cellular coverage to ground users \cite{loonlibrary}.

%In 2017 large areas of Peru were hit with severe flooding that damaged the terrestrial infrastructures, causing the interruption of communications.
%The solution adopted was to use several Loon balloons 
%Furthermore, in 2020, Loon was deployed to bring Internet connection to unserved regions of Kenya, launching the world first internet-via-balloon service.
%hile Loon's project has been currently completed, there are many other active projects that are achieving interesting results, such as Airbus’s Zephyr~\cite{zephyr} and Sceye~\cite{sceye}.

Besides providing connectivity, \glspl{ntn} can also host edge servers for processing, caching, and/or storing data generated from power-constrained \gls{iot} devices, thereby supporting the transition of remote areas towards smart cities~\cite{nguyen20216g}. Similarly, air/space-borne platforms can support \gls{vec}, where \glspl{gv} offload computationally-intensive autonomous driving tasks like object detection and recognition, tracking/trajectory prediction, and/or semantic segmentation~\cite{liu2021vehicular}. Processing these tasks onboard certain vehicles may be infeasible due to the limited autonomy and to the small capacity of their batteries. While in urban areas \glspl{gv} can delegate the burden of data processing to roadside units~\cite{v2x_offloading}, NTNs may represent a valid alternative to serve computational requests in poorly connected rural areas.
%Therefore, energy and resource constrained cars could offload intensive tasks and caching~\cite{ren2021caching} to powerful \gls{nt} servers, connected via \gls{ntn}, in order to process their \gls{v2x} application~\cite{shinde2021towards}.
One of the main requirements of VEC systems is the support for real-time latency to guarantee safe autonomous driving as specified by the standards~\cite{velez20205g}.
% Satellites fly at high altitudes, which would incur large round-trip times, likely beyond real-time requirements even considering only air latency.
% UAVs, instead, offer limited computational capacity in view of flight time limitations, and seem more helpful to complement terrestrial coverage when cellular infrastructures are overloaded.
While satellites incur large round-trip times, UAVs offer limited computational capacity and short-lived connectivity~\cite{ke2021edge}. 
Instead, HAPs provide large coverage, and are deployed in the stratosphere at an altitude of 20~km, which helps reduce the propagation delay to less than 1~ms.
The effectiveness of \glspl{hap} in \gls{mec} networks was studied in \cite{hap_offloading}, where the authors addressed the problems of partial offloading and bandwidth allocation in remote areas without the coverage of ground infrastructure.
However the benefit of \glspl{hap} for \gls{vec} is still an open question, that we will analyze in this work.
% Based on the above introduction, in this paper we present an optimization problem for HAP-assisted VEC in a rural scenario in which several \glspl{gv} generate perception data in the form of video frames. For each frame, the GV can decide whether to process it onboard, or offload it to the \gls{hap}.

Based on the above introduction, in this paper we present an optimization problem for real-time VEC in a rural scenario, in which multiple \glspl{gv} can decide whether to process perception data (generated in the form of video frames) onboard or offload them to a \gls{hap}.
% Notably, we address the research question: \emph{``Is it practical to offload self-driving-related tasks to HAPs?"}
We consider wireless transmissions in the \gls{mmwave} band~\cite{wang2020potential}
%and model the arrival process of computing tasks according to a Poisson process with a rate proportional to the number of GVs.
%Given the need for \gls{vec} systems to support real-time processing, in this paper, we 
With respect to our previous work~\cite{traspadini2022uavhapassisted}, we develop a more accurate model to characterize both \glspl{gv}' and \gls{hap}'s queuing systems, and determine a closed-form expression for the average waiting time experienced by computing tasks. 
We show that it is convenient to offload some VEC processing tasks from GVs to the HAP under common assumptions for the processing capacity and load of the nodes. 
%In the remainder of this paper we first introduce our system model in~\cref{sec:system_model}, then we describe our optimization problem  in~\cref{sec:optimizing_offloading_for_ntns}, while in~\cref{sec:performace} we show our results. Conclusions are summarized in~\cref{sec:conclusions_and_future_works}.

\section{System Model}
\label{sec:system_model}

%In this section, we present our research problem, as well as the delay, channel and queuing models.

\subsection{Problem Formulation}
\label{sub:formulation}

We analyze a vast remote region where $n$ \glspl{gv} are distributed within an \gls{aoi} of size~$A$.
Each \Gls{gv}~generates sensory perceptions in the form of video frames of size $n_{\rm UL}$ according to a Poisson process with arrival rate~$r$, as considered in some reference papers to model autonomous driving data~\cite{lundgren2014vehicle}.
Each frame requires a constant computational load $C$ (e.g., for object detection~\cite{rossi2021role}), which has to be executed fast enough to provide real-time services, i.e., at least at the frame rate of the sensor.

%The \gls{aoi} is a remote region with no ground infrastructures to provide GVs with computing servers.
We consider that the \glspl{gv} can offload a subset $\eta$ of their computational load to a \gls{hap} providing \gls{vec} functionalities. 
According to the split property of the Poisson processes, we can distinguish two independent Poisson processes, namely for frames that are offloaded to the HAP at a rate~$\eta r$, and for frames that are processed onboard at a rate~$(1-\eta) r$.
On one side, local processing may involve long delays, given the limited computational capacity $C_{\rm GV}$ of low-budget car models.
On the other side, the HAP can count on a higher computing power $C_{\rm HAP}$ than available onboard a vehicle, i.e., $C_{\rm HAP}\geq C_{\rm GV}$, thus processing data faster, at the expense of a non-negligible communication delay for offloading data to the HAP server and for the processed data to be returned to the end users.
We will investigate the optimized selection of $\eta \in [0, 1]$.

%A \gls{gv}, indeed, can offer only limited computation capacity $C_{\rm GV}$~GFLOPs, therefore it might not be able to manage the whole perception rate satisfying the latency constraint.
%A \gls{hap} server, instead, provides a higher computational capacity $C_{\rm HAP}$~GFLOPs, but it has to serve the traffic of all the \glspl{gv} in the \gls{aoi}. Furthermore transferring information over a wireless link back and forth to the \gls{hap} introduces additional non-negligible delays.

\subsection{Delay Model} % (fold)
\label{sub:delay_model}
This section introduces our delay model for each data frame for both onboard and \gls{hap}-assisted processing.

\paragraph{Onboard processing}
Each \gls{gv} is modeled as an M/D/$1$ queue, and the average delay is equal to
\begin{equation}
	\bar{t}_\text{GV} = \bar{W}_q{({\text{M/D/}1})}+ \Big({C}/{C_\text{GV}}\Big),
	\label{eq:t_lp}
\end{equation}
where $\bar{W}_q{({\text{M/D/}1})}$ is the average waiting time (see Sec.~\ref{sub:queuing_model}), while ${C}/{C_\text{GV}}$  is the frame processing time onboard the GV.

\paragraph{HAP-assisted processing}
Since the HAP has less severe space and energy constraints than a GV, it is modeled as an M/D/$c$ queue, where $c$ servers can process up to $c$ frames in parallel. 
The average delay for processing tasks can be expressed as
\begin{equation}
\bar{t}_\text{HAP} = 2\tau_p + t_{\rm UL}+t_{\rm DL} +  \bar{W}_q{({\text{M/D/}c})}+ \Big({C}/{C_\text{HAP}}\Big),
\label{eq:t_hap}
\end{equation}
where $\bar{W}_q{({\text{M/D/}c})}$ is the average waiting time (see Sec.~\ref{sub:queuing_model}), $\tau_p=d/c_l$ is the propagation delay (where $d$ is the distance between the generic GV and the HAP, and $c_l$ is the speed of light), and $t_{\rm UL}$ and $t_{\rm DL}$ are the times to transmit data to and from the HAP, respectively  (see Sec.~\ref{sub:channel}).

\subsection{HAP-to-Ground Channel Model} 
\label{sub:channel}
According to the 3GPP specifications~\cite{38821}, HAPs may operate through highly directional links at \glspl{mmwave}, where the large bandwidth available at these frequencies offers the potential for ultra-fast connectivity~\cite{wang2020potential}.
With the assumption of an interference-free environment considering transmissions over orthogonal bands, 
the \gls{snr} between transmitter $i$ and receiver $j$ can be calculated as:
%\begin{equation}
%	\label{eq:snr}
%	\gamma_{ij} = \text{EIRP}_i + ({G_j}/{T}) - \text{PL}_{ij} - k - B,
%\end{equation}

\begin{equation}
\label{eq:snr}
\gamma_{ij} = \frac{\text{EIRP}_i \cdot ({G}/{T})_j}{\text{PL}_{ij} \cdot k \cdot B}
\end{equation}

where $\text{EIRP}_i$ is the effective isotropic radiated power, $({G}/{T})_j$ is the receiver antenna-gain-to-noise-temperature, PL is the path loss, $k$ is the Boltzmann constant, and $B$ is the bandwidth. The path loss depends on the frequency (here \glspl{mmwave}) and on the distance between $i$ and $j$, and accounts for additional atmospheric attenuations as described in~\cite{giordani2020satellite}.

Based on the SNR in Eq.~\eqref{eq:snr}, we introduce the median ergodic capacity $R$ between the generic GV and the HAP as
	$R=B\log_2(1+\gamma_{ij})$.
Then, the transmission times $t_{\rm UL}$ and $t_{\rm DL} $ in Eq.~\eqref{eq:t_hap}, for each data frame, are given by
	$t_\ell = n_{\ell}/R_{\ell}, \quad\ell\in\{\text{UL,DL}\},$
where 
%$n$ is the number of GVs in the \gls{aoi}, 
$n_\ell$ is the size of the transmitted data.

\subsection{Queuing Model}
\label{sub:queuing_model}
Sensory frames are generated at rate $r$, and are queued and eventually processed at the GVs and HAP according to a Poisson process of rate $\lambda$, equal to $(1-\eta)r$ and $\eta r n$, respectively; we follow a \gls{fcfs} discipline. 
Notably, GVs operate via an M/D/$1$ queue, while HAPs via an M/D/$c$ queue. %At time $t$, a general M/D/$c$ queue (with arrival rate $\lambda$ and service rate $\mu$) provides a constant service time of duration $1/\mu$.
%At time $t+(1/\mu)$, the queue consists of both old frames already in queue at time $t$ and not yet processed, and new frames generated in $[t,t+(1/\mu)]$.
Following the analysis made in~\cite{tijms2003Afirstcourse}, the  M/D/$c$ state probability $p_j$ for state $j$, i.e., the probability that the queue at the HAP has $j$ frames, can be computed by observing that any frame in the system at time $t+1/\mu$ (where $1/\mu$ is the constant processing time) is either waiting in the queue at time $t$ or has arrived in $[t,t+1/\mu]$.
Thus, given that the number of arrivals in an interval of duration $1/\mu$ is Poisson distributed with mean $\lambda/\mu$, the state probability $p_j$ is:
\begin{equation}
	p_j = e^{-G}\frac{G^j}{j!}\sum_{k = 0}^{c}p_k + e^{-G} \sum_{k = c+1}^{c+j}p_k\frac{G^{j-k+c}}{(j-k+c)!},
	\label{eq:p_j}
\end{equation}
where $G=\lambda/\mu$ is the offered traffic, and $\mu$ is the service rate at the HAP and GVs, i.e., $C_{\rm HAP}/C$ and $C_{\rm GV}/C$, respectively.
Eq.~\eqref{eq:p_j}, combined with the normalization condition $\left(\sum_{j=0}^{+\infty}p_j = 1\right)$, generates an infinite system of linear equations that describe the state probabilities of the queue. We adopted the geometric tail approximation proposed in~\cite{tijms2003Afirstcourse} to find a closed-form expression for this system of equations. Therefore, we can write 
\begin{equation}
p_j=p_M \tau^{-(j-M)}, \quad j\geq M,
\label{eq:p_j_approx}
\end{equation}
where $\tau$ is a constant depending on the server utilization, and $M$ is a threshold state; we refer the interested reader to ~\cite{tijms2003Afirstcourse} to clarify how to compute these values.
Based on Eq.~\eqref{eq:p_j_approx}, the infinite system in Eq.~\eqref{eq:p_j} can be reduced to a finite system of $M+1$ linear equations that can be solved easily.
\begin{lem}
Based on Eq.~\eqref{eq:p_j_approx}, an empirical closed-form expression for the average waiting time of an M/D/$c$ queue with arrival rate $\lambda$ and service rate $\mu$ is given by
\begin{equation}\label{eq:Wq1}
\bar{W}_q{({\text{M/D/}c})}=\frac{\sum_{k=c}^{M-1} p_k (k-c) + p_M \left[\frac{M-c+\frac{1}{1-\frac{1}{\tau}}-1}{1-\frac{1}{\tau}}\right]}{\lambda},
\end{equation}
where a relatively small value of $M$ can already provide an accurate approximation of $\bar{W}_q{({\text{M/D/}c})}$.
\end{lem}
\renewcommand\qedsymbol{$\blacksquare$}
\begin{proof}
See Appendix.
\end{proof}

\section{Optimal Offloading for HAP-Assisted VEC}
\label{sec:optimizing_offloading_for_ntns}

In our scenario, we require that autonomous driving tasks, especially object detection on the video frames, be executed at least at the same rate at which new frames are generated, i.e., $t_\text{max}=1/r$. We model this aspect through the probability of real-time service $P_{\rm RT}$, given by
\begin{equation}
P_{\rm RT}(\eta) = \eta P(t_\text{HAP}\leq t_\text{max}) + (1-\eta) P(t_\text{GV}\leq t_\text{max}),
\label{eq:p_rt}
\end{equation}
where $t_\text{max}$ is the maximum tolerable latency, and $t_\text{GV}$ and $t_\text{HAP}$ have been defined in Eqs.~\eqref{eq:t_lp} and~\eqref{eq:t_hap}, respectively.
Given the rural scenario, we assume that all \glspl{gv} experience the same channel condition.  Thus, we assume they follow the same policy, and offload data frames with the same probability $\eta$ (\emph{shared offloading factor}). 
%The objective of our optimization problem is then to select $\eta$ so as $P_{\rm RT}(\eta)$  is maximized.
The value of $P(t_\text{HAP}\leq t_\text{max})$ in Eq.~\eqref{eq:p_rt} can be expressed by computing the maximum number of frames in the system that ensure real-time processing, that~is
\begin{equation}
f^\text{HAP}_\text{max} = c \left \lfloor \frac{t_\text{max}-t_{\rm UL}-t_{\rm DL}-2\tau_p}{C/C_{\rm HAP}} \right \rfloor.
\end{equation}
However, real-time processing is also possible when new offloading requests find the queue at full capacity, but some processing tasks are already in service and will leave the system soon.
We define as $\Delta^\text{HAP}_t$ the percentage of in-service tasks that ensure real-time processing even if more than $f^\text{HAP}_\text{max}$ requests are in the system. We have
\begin{equation}
\Delta^\text{HAP}_t = \Gamma\left(\frac{t_\text{max}-t_{\rm UL}-t_{\rm DL}-2\tau_p}{C/C_{\rm HAP}}\right),
\end{equation}
where $ \rm \Gamma(x)=x-\lfloor x \rfloor$.
Similarly, we can solve $P(t_\text{GV}\leq t_\text{max})$ in Eq.~\eqref{eq:p_rt} by introducing $f^\text{GV}_\text{max}$ and $\Delta^\text{GV}_t $ for the \gls{gv}:
\begin{equation}
f^\text{GV}_\text{max} = \left \lfloor \frac{t_\text{max}}{C/C_{\rm GV}} \right \rfloor, \quad \Delta^\text{GV}_t = \Gamma\left(\frac{t_\text{max}}{C/C_{\rm GV}} \right). 
\end{equation}
Eq.~\eqref{eq:p_rt} depends on the queuing state probability of both \glspl{gv} and \gls{hap}, based on the model in~\cref{sub:queuing_model}.
Using the Poisson Arrivals See Time Averages (PASTA) property, and the probability distribution described in Eqs.~\eqref{eq:p_j} and \eqref{eq:p_j_approx}, the real-time probability for the HAP can be expressed as
\begin{equation}\label{uno}
 \begin{aligned}
 &P(t_\text{HAP}\leq t_\text{max}) = \\
 &\sum_{k = 1}^{c} p_\text{bin}(k,c,\Delta^\text{HAP}_t) \sum_{j=0}^{k-1}p_{f^\text{HAP}_\text{max}+j}+\sum_{i=0}^{f^\text{HAP}_\text{max}-1} p_i,
 \end{aligned}
\end{equation}
where $p_\text{bin}(k,c,\Delta^\text{HAP}_t) = \binom{c}{k} \left(\Delta^\text{HAP}_t\right)^{k} \left(1-\Delta^\text{HAP}_t\right)^{c-k}$ describes the probability that $k$ offloading requests have been processed by at least a fraction $\Delta^\text{HAP}_t$, and probabilities $p_i$ are as in Eq.~\eqref{eq:p_j}. 
Similarly, for the GV we~have
\begin{equation}\label{due}
P(t_\text{GV}\leq t_\text{max}) = \sum_{i=0}^{f^\text{GV}_\text{max}-1} p_i + \Delta^\text{GV}_t p_{f^\text{GV}_\text{max}}.
\end{equation}
The objective of our optimization problem is to choose the optimal offloading factor $\eta$, i.e., $\eta^*$, that maximizes $P_{\rm RT}$, i.e.,
\begin{subequations}
\label{eq:opt_problem_draft}
\begin{alignat}{2}
&\argmax_{\eta} &\quad& P_{\rm RT}(\eta), \\
&\text{subject to} & & G_{\rm HAP}< c, \: \: G_{\rm GV}< 1 \label{eq:queue_constraint}\\
&  & & \eta \in [0,1],
\end{alignat}
\end{subequations}
where $G_{\rm HAP} = \eta r C n/C_{\rm HAP}$ and $G_{\rm GV} = (1-\eta)r C/C_{\rm GV}$ are the offered traffic at the \gls{hap} and the \glspl{gv}, respectively, while \cref{eq:queue_constraint} %are necessary to keep both \glspl{gv} and \gls{hap} queues stable, and
%Besides, constraints in \cref{eq:queue_constraint,eq:queue_constraint_2} only 
sets a constraint on $\eta$ to keep the system stable.
Therefore, the optimization problem can be rewritten~as
\begin{subequations}
\label{eq:opt_problem}
\begin{alignat}{2}
&\argmax_{\eta} &\quad& P_{\rm RT}(\eta), \\
&\text{subject to} & & \eta \in [\rm \eta_{min},\eta_{max}],
\end{alignat}
\end{subequations}
where $ \eta_{\rm min} = \max(0,1-C_{\rm GV} / (rC))$ and $ \eta_{\rm max} = \min(1,c C_{\rm HAP} / (r n C))$. %are the stability limits of M/D/1 and M/D/$c$ queues combined with the possible ranges of $\eta$. 
Notably, $C_{\rm GV}$ ($C_{\rm HAP}$) set a limit to the lower (upper) bound of $\eta$. 
Given that $P_{\rm RT}(\eta)$ is a continuous function defined on a closed interval $[\eta_{\rm min},\,\eta_{\rm max}]	$, there exists a solution to the optimization problem in \cref{eq:opt_problem} according to Weierstrass' theorem.
Then, the problem is solved numerically using the Brent solver~\cite{gegenfurtner1992praxis}.

%The solution of this optimization problem provides also some valuable insights on the reliability and safety that a certain scenario can guarantee to \glspl{gv}, for instance defining a minimum value for $P_{\rm RT}$.
We can analyze the 
%average latency for executing autonomous driving tasks,  defined as the 
average time it takes for a GV, which may offload computational tasks to the HAP, to process each video frame.
Since, on average, each GV processes a frame in $\bar{t}_\text{GV}$, and the latency for an offloaded frame is $\bar{t}_\text{HAP}$, then the average latency is evaluated as
\begin{equation}
\label{eq:avg}
\bar{t}(\eta) = \eta^* \bar{t}_\text{HAP} + (1-\eta^*) \bar{t}_\text{GV},
\end{equation}
where $\eta^*$ is the optimal offloading probability from Eq.~\eqref{eq:opt_problem}.

\section{Performance evaluation}
\label{sec:performace}
In Sec.~\ref{sub:simulation_setup_and_parameters} below, we introduce our simulator and its parameters, while numerical results are provided in Sec.~\ref{sub:numerical_results}.

\subsection{Simulation Setup and Parameters}
	\label{sub:simulation_setup_and_parameters}

%	for different offloading strategies: the optimal offloading factor 
%	(Optimal Policy), a constant offloading factor equal to 30~$\%$ of the overall amount of perceptions (Fixed Policy) and the not assisted scenario (Fully Local).
%	The Fixed Policy can be applied when the vehicles have not a comprehensive knowledge of the scenario (communication channel and number of users that require offloading), thus they cannot evaluate the optimal policy.
%	We decide to set the offloading factor for the Fixed Policy to only 0.3 in order to keep the users more cautions.
%	A higher offloading factor would be more prone to instability and would have a higher dependence on the number of vehicles in the system.

\begin{figure*}[t!]
	\begin{subfigure}[b]{\linewidth}
		\centering
		\setlength\fwidth{\columnwidth}
		% This file was created by matlab2tikz.
%
%The latest updates can be retrieved from
%  http://www.mathworks.com/matlabcentral/fileexchange/22022-matlab2tikz-matlab2tikz
%where you can also make suggestions and rate matlab2tikz.
%

\definecolor{black25}{RGB}{25,25,25}
\definecolor{chocolate2238331}{RGB}{223,83,31}
\definecolor{darkslategray38}{RGB}{38,38,38}
\definecolor{darkslategray66}{RGB}{66,66,66}
\definecolor{lightgray204}{RGB}{204,204,204}
\definecolor{limegreen3122331}{RGB}{31,223,31}
\definecolor{royalblue58143226}{RGB}{58,143,226}
\definecolor{silver}{RGB}{192,192,192}
\definecolor{steelblue76114176}{RGB}{76,114,176}

\begin{tikzpicture}
\pgfplotsset{every tick label/.append style={font=\scriptsize}}

\pgfplotsset{compat=1.11,
	/pgfplots/ybar legend/.style={
		/pgfplots/legend image code/.code={%
			\draw[##1,/tikz/.cd,yshift=-0.25em]
			(0cm,0cm) rectangle (10pt,0.6em);},
	},
}

\begin{axis}[%
width=0,
height=0,
at={(0,0)},
scale only axis,
xmin=0,
xmax=0,
xtick={},
ymin=0,
ymax=0,
ytick={},
axis background/.style={fill=white},
legend style={legend cell align=left,
              align=center,
              draw=white!15!black,
              at={(0, 0)},
              anchor=center,
              /tikz/every even column/.append style={column sep=1em}},
legend columns=5,
]
\addplot[ybar,ybar legend,draw=black,fill=chocolate2238331,line width=0.08pt,postaction={
	pattern=dots}]
table[row sep=crcr]{%
	0	0\\
};
\addlegendentry{Fully local}

%\addplot[ybar,ybar legend,draw=black,fill=limegreen3122331,line width=0.08pt]
%  table[row sep=crcr]{%
%	0	0\\
%};
%\addlegendentry{Fixed Policy}

\addplot[ybar legend,ybar,draw=black,fill=royalblue58143226,line width=0.08pt]
  table[row sep=crcr]{%
	0	0\\
};
\addlegendentry{HAP-assisted VEC}

\addplot[ybar,ybar legend,draw=black,fill=white,line width=0.08pt,postaction={
	pattern=north east lines
}]
table[row sep=crcr]{%
	0	0\\
};
\addlegendentry{Instability}

\addplot [semithick, steelblue76114176,xshift = -0.15cm, mark=diamond*, mark size=4, mark options={solid,fill=silver,draw=black25}, only marks]
table[row sep=crcr]{%
	-0	0\\
};
\addlegendentry{Offloading factor}

\addplot [color=black, dashed, line width=1.5pt,xshift = -0.15cm]
table[row sep=crcr]{%
	0	0\\
};
%\addlegendentry{Maximum tolerable latency ($t_{\rm max}=1/r$)}
\addlegendentry{$t_{\rm max}=1/r$}

\end{axis}
\end{tikzpicture}%
	\end{subfigure}
	%\vskip 0.5cm
	%\centering
	\subfloat[][$C_{\rm GV}=$ 800~GFLOPS.]
	{
		\label{fig:density}
		% This file was created with tikzplotlib v0.10.1.
\begin{tikzpicture}[scale=1]

\definecolor{black25}{RGB}{25,25,25}
\definecolor{darkslategray38}{RGB}{38,38,38}
\definecolor{darkslategray66}{RGB}{66,66,66}
\definecolor{lightgray204}{RGB}{204,204,204}
\definecolor{red}{RGB}{223,83,31}
\definecolor{royalblue58143226}{RGB}{58,143,226}
\definecolor{silver}{RGB}{192,192,192}
\definecolor{steelblue76114176}{RGB}{76,114,176}

\pgfplotsset{
	tick label style={font=\scriptsize},
	label style={font=\scriptsize},
	legend  style={font=\scriptsize}
}

\begin{axis}[
width = \textwidth/2.3,
height = 4cm,
axis line style={lightgray204},
tick align=outside,
unbounded coords=jump,
x grid style={lightgray204},
xlabel=\textcolor{darkslategray38}{Number of GVs $n$},
xmajorticks=true,
xmin=-0.5, xmax=3.5,
xtick style={color=darkslategray38},
xtick={0,1,2,3},
xticklabels={50,90,150,200},
y grid style={lightgray204},
ylabel=\textcolor{darkslategray38}{Average latency $\bar{t}(\eta)$ [s]},
ymajorgrids,
ymin=0, ymax=0.3,
ytick pos=left,
ytick style={color=darkslategray38}
]
\draw[draw=black,fill=royalblue58143226,line width=0.08pt] (axis cs:-0.35,0) rectangle (axis cs:-0.05,0.0382611782736381);
%\addlegendimage{ybar,ybar legend,draw=black,fill=royalblue58143226,line width=0.08pt}
%\addlegendentry{opt}

\draw[draw=black,fill=royalblue58143226,line width=0.08pt] (axis cs:0.65,0) rectangle (axis cs:0.95,0.0945684020609371);
\draw[draw=black,fill=royalblue58143226,line width=0.08pt] (axis cs:1.65,0) rectangle (axis cs:1.95,0.130307882563182);
\draw[draw=black,fill=royalblue58143226,line width=0.08pt] (axis cs:2.65,0) rectangle (axis cs:2.95,0.142172693490478);
\draw[draw=black,fill=red,line width=0.08pt,postaction={
	pattern=dots}] (axis cs:0.05,0) rectangle (axis cs:0.35,0.18728853877192);

%\addlegendimage{ybar,ybar legend,draw=black,fill=red,line width=0.08pt}
%\addlegendentry{Fully-Local}

\draw[draw=black,fill=red,line width=0.08pt,postaction={
	pattern=dots}] (axis cs:1.05,0) rectangle (axis cs:1.35,0.18728853877192);
\draw[draw=black,fill=red,line width=0.08pt,postaction={
	pattern=dots}] (axis cs:2.05,0) rectangle (axis cs:2.35,0.18728853877192);
\draw[draw=black,fill=red,line width=0.08pt,postaction={
	pattern=dots}] (axis cs:3.05,0) rectangle (axis cs:3.35,0.18728853877192);
\addplot [line width=1.08pt, darkslategray66]
table {%
	-0.2 nan
	-0.2 nan
};
\addplot [line width=1.08pt, darkslategray66]
table {%
	0.8 nan
	0.8 nan
};
\addplot [line width=1.08pt, darkslategray66]
table {%
	1.8 nan
	1.8 nan
};
\addplot [line width=1.08pt, darkslategray66]
table {%
	2.8 nan
	2.8 nan
};
\addplot [line width=1.08pt, darkslategray66]
table {%
	0.2 nan
	0.2 nan
};
\addplot [line width=1.08pt, darkslategray66]
table {%
	1.2 nan
	1.2 nan
};
\addplot [line width=1.08pt, darkslategray66]
table {%
	2.2 nan
	2.2 nan
};
\addplot [line width=1.08pt, darkslategray66]
table {%
	3.2 nan
	3.2 nan
};
\addplot [very thick, black, dashed]
table {%
	-10 0.1
	-9 0.1
	-8 0.1
	-7 0.1
	-6 0.1
	-5 0.1
	-4 0.1
	-3 0.1
	-2 0.1
	-1 0.1
	0 0.1
	1 0.1
	2 0.1
	3 0.1
	4 0.1
	5 0.1
	6 0.1
	7 0.1
	8 0.1
	9 0.1
};
\end{axis}

\begin{axis}[
width = \textwidth/2.3,
height = 4cm,
axis line style={lightgray204},
tick align=outside,
x grid style={lightgray204},
xmajorticks=false,
xmin=-0.5, xmax=3.5,
xtick style={color=darkslategray38},
y grid style={lightgray204},
ylabel=\textcolor{darkslategray38}{Optimal offloading factor \(\displaystyle \eta^*\)},
ylabel style={font=\scriptsize},
ymin=-0.05, ymax=1.05,
ytick pos=right,
ytick style={color=darkslategray38},
yticklabel style={anchor=west},
]
\addplot [semithick, steelblue76114176, mark=diamond*, mark size=5, mark options={solid,fill=silver,draw=black25}, only marks]
table {%
-0.2 1
% 0.2 0
0.8 0.690723751495036
% 1.2 0
1.8 0.256670679311554
% 2.2 0
2.8 0.256670679311554
% 3.2 0
};
\end{axis}

\end{tikzpicture}
	}
	\subfloat[][$n= $ 90~GVs.]
	{
		\label{fig:capacity}
		% This file was created with tikzplotlib v0.10.1.
\begin{tikzpicture}[scale=1]

\definecolor{black25}{RGB}{25,25,25}
\definecolor{darkslategray38}{RGB}{38,38,38}
\definecolor{darkslategray66}{RGB}{66,66,66}
\definecolor{lightgray204}{RGB}{204,204,204}
\definecolor{red}{RGB}{223,83,31}
\definecolor{royalblue58143226}{RGB}{58,143,226}
\definecolor{silver}{RGB}{192,192,192}
\definecolor{steelblue76114176}{RGB}{76,114,176}

\pgfplotsset{
	tick label style={font=\scriptsize},
	label style={font=\scriptsize},
	legend  style={font=\scriptsize}
}

\begin{axis}[
width = \textwidth/2.35,
height = 4cm,
axis y line=left,
ylabel style={font=\scriptsize},
axis line style={lightgray204},
tick align=outside,
unbounded coords=jump,
x grid style={lightgray204},
xlabel=\textcolor{darkslategray38}{Computational capacity \(\displaystyle C_{\rm GV}\) [GFLOPS]},
xmajorticks=true,
xmin=-0.5, xmax=3.5,
xtick style={color=darkslategray38},
xtick={0,1,2,3},
xticklabels={200,600,800,1000},
y grid style={lightgray204},
ylabel=\textcolor{darkslategray38}{Average latency $\bar{t}(\eta)$ [s]},
ymajorgrids,
ymin=0, ymax=0.5,
ytick pos=left,
ytick style={color=darkslategray38}
]
\draw[draw=black,fill=royalblue58143226,line width=0.08pt] (axis cs:-0.35,0) rectangle (axis cs:-0.05,0.174982374084067);
%\addlegendimage{ybar,ybar legend,draw=black,fill=royalblue58143226,line width=0.08pt}
%\addlegendentry{opt}

\draw[draw=black,fill=royalblue58143226,line width=0.08pt] (axis cs:0.65,0) rectangle (axis cs:0.95,0.103625447489147);
\draw[draw=black,fill=royalblue58143226,line width=0.08pt] (axis cs:1.65,0) rectangle (axis cs:1.95,0.0945684020609371);
\draw[draw=black,fill=royalblue58143226,line width=0.08pt] (axis cs:2.65,0) rectangle (axis cs:2.95,0.0874793166717147);
%\addlegendimage{ybar,ybar legend,draw=black,fill=red,line width=0.08pt,postaction={pattern=north east lines}}
%\addlegendentry{Fully-Local}

%\draw[draw=black,fill=red,line width=0.08pt,postaction={pattern=north east lines}] (axis cs:0.05,0) rectangle (axis cs:0.35,10);
%\draw[draw=black,fill=red,line width=0.08pt,postaction={pattern=north east lines}] (axis cs:1.05,0) rectangle (axis cs:1.35,10);
%\draw[draw=black,fill=red,line width=0.08pt,postaction={pattern=dots}] (axis cs:2.05,0) rectangle (axis cs:2.35,0.18728853877192);
%\draw[draw=black,fill=red,line width=0.08pt,postaction={
%	pattern=dots}] (axis cs:3.05,0) rectangle (axis cs:3.35,0.104928910274621);

\draw[draw=black,fill=red, postaction={pattern=north east lines},line width=0.08pt] (axis cs:0.05,0) rectangle (axis cs:0.35,0.42);
\draw[draw=black,fill=red, postaction={pattern=north east lines}, path fading =north,line width=0.08pt] (axis cs:0.05,0.44) rectangle (axis cs:0.35,0.5);
\draw[draw=black,path fading =north, postaction={pattern=north east lines},line width=0.2pt, dashed] (axis cs:0.05,0) rectangle (axis cs:0.35,0.5);
\addplot[black] coordinates {(0,0.42) (0.4,0.42)};
\addplot[black] coordinates {(0,0.44) (0.4,0.44)};

\draw[draw=black,fill=red, postaction={pattern=north east lines},line width=0.08pt] (axis cs:1.05,0) rectangle (axis cs:1.35,0.42);
\draw[draw=black,fill=red, postaction={pattern=north east lines}, path fading =north,line width=0.08pt] (axis cs:1.05,0.44) rectangle (axis cs:1.35,0.5);
\draw[draw=black,path fading =north, postaction={pattern=north east lines},line width=0.2pt, dashed] (axis cs:1.05,0) rectangle (axis cs:1.35,0.5);
\addplot[black] coordinates {(1,0.42) (1.4,0.42)};
\addplot[black] coordinates {(1,0.44) (1.4,0.44)};

\addplot [line width=1.08pt, darkslategray66]
table {%
-0.2 nan
-0.2 nan
};
\addplot [line width=1.08pt, darkslategray66]
table {%
0.8 nan
0.8 nan
};
\addplot [line width=1.08pt, darkslategray66]
table {%
1.8 nan
1.8 nan
};
\addplot [line width=1.08pt, darkslategray66]
table {%
2.8 nan
2.8 nan
};
\addplot [line width=1.08pt, darkslategray66]
table {%
0.2 nan
0.2 nan
};
\addplot [line width=1.08pt, darkslategray66]
table {%
1.2 nan
1.2 nan
};
\addplot [line width=1.08pt, darkslategray66]
table {%
2.2 nan
2.2 nan
};
\addplot [line width=1.08pt, darkslategray66]
table {%
3.2 nan
3.2 nan
};
\addplot [very thick, black, dashed]
table {%
-10 0.1
-9 0.1
-8 0.1
-7 0.1
-6 0.1
-5 0.1
-4 0.1
-3 0.1
-2 0.1
-1 0.1
0 0.1
1 0.1
2 0.1
3 0.1
4 0.1
5 0.1
6 0.1
7 0.1
8 0.1
9 0.1
};
\end{axis}

\begin{axis}[
width = \textwidth/2.35,
height = 4cm,
axis line style={lightgray204},
tick align=outside,
x grid style={lightgray204},
xmajorticks=false,
xmin=-0.5, xmax=3.5,
xtick style={color=darkslategray38},
y grid style={lightgray204},
ylabel=\textcolor{darkslategray38}{Optimal offloading factor $\eta^*$},
ylabel style={font=\scriptsize},
ymin=-0.05, ymax=1.05,
ytick pos=right,
ytick style={color=darkslategray38},
yticklabel style={anchor=west}
]
\addplot [semithick, steelblue76114176, mark=diamond*, mark size=5, mark options={solid,fill=silver,draw=black25}, only marks]
table {%
-0.2 0.809641574171659
% 0.2 0
0.8 0.735522646995212
% 1.2 0
1.8 0.690723751495036
% 2.2 0
2.8 0.641089214436178
% 3.2 0
};
\end{axis}

\end{tikzpicture}
	}
    \caption{Optimal offloading factor $\eta^*$ (right axis) and average latency $\bar{t}(\eta)$ (left axis) vs. $n$ and $C_{\rm GV}$, for $r=10$ fps and $n_\text{UL}=3$~Mb. Striped bars are interrupted to represent the case of unstable queues where the latency increases indefinitely in the long term.\vspace{-0.4cm}}
	\label{fig:analysis}
\end{figure*}
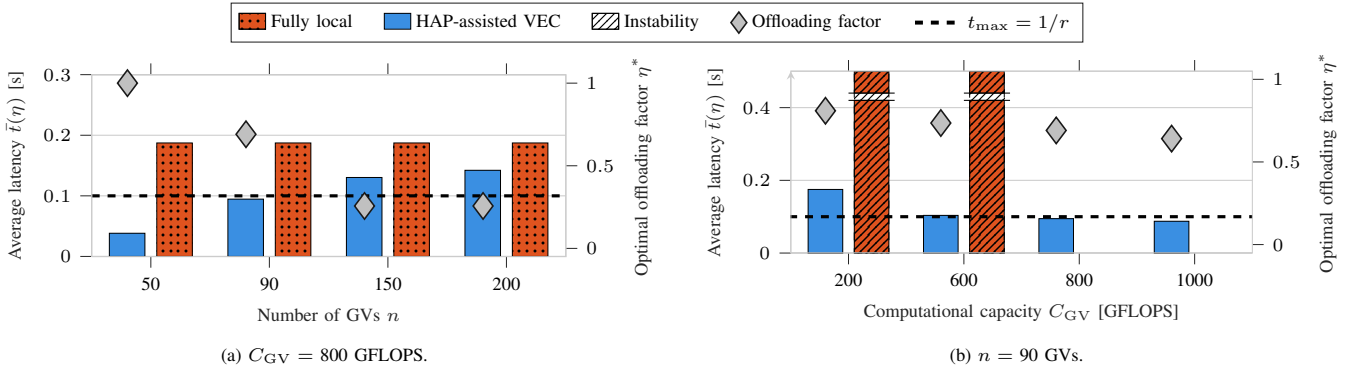

\begin{figure*}[t!]
	\begin{subfigure}[b]{\linewidth}
		\centering
		\setlength\fwidth{\columnwidth}
		% This file was created by matlab2tikz.
%
%The latest updates can be retrieved from
%  http://www.mathworks.com/matlabcentral/fileexchange/22022-matlab2tikz-matlab2tikz
%where you can also make suggestions and rate matlab2tikz.
%

\definecolor{black25}{RGB}{25,25,25}
\definecolor{chocolate2238331}{RGB}{223,83,31}
\definecolor{darkslategray38}{RGB}{38,38,38}
\definecolor{darkslategray66}{RGB}{66,66,66}
\definecolor{lightgray204}{RGB}{204,204,204}
\definecolor{limegreen3122331}{RGB}{31,223,31}
\definecolor{royalblue58143226}{RGB}{58,143,226}
\definecolor{silver}{RGB}{192,192,192}
\definecolor{steelblue76114176}{RGB}{76,114,176}

\definecolor{gainsboro202218233}{RGB}{202,218,233}
\definecolor{cornflowerblue119170200}{RGB}{119,170,200}
\definecolor{steelblue50110161}{RGB}{50,110,161}

\begin{tikzpicture}

\draw [decorate,decoration={brace,mirror,amplitude=7pt},xshift=0pt,yshift=-0.2cm](5,0.5) -- (-7,0.5) node[black,midway,above,xshift=0cm,yshift=0.2cm] 
{\scriptsize HAP-assisted VEC};

\pgfplotsset{every tick label/.append style={font=\scriptsize}}

\pgfplotsset{compat=1.11,
	/pgfplots/ybar legend/.style={
		/pgfplots/legend image code/.code={%
			\draw[##1,/tikz/.cd,yshift=-0.25em]
			(0cm,0cm) rectangle (10pt,0.6em);},
	},
}

\begin{axis}[%
width=0,
height=0,
at={(0,0)},
scale only axis,
xmin=0,
xmax=0,
xtick={},
ymin=0,
ymax=0,
ytick={},
axis background/.style={fill=white},
legend style={legend cell align=left,
              align=center,
              draw=white!15!black,
              at={(0, 0)},
              anchor=center,
              %/tikz/every even column/.append style={column sep=1em},
              /tikz/column 2/.style={column sep=1em},
              /tikz/column 4/.style={column sep=1em},
              /tikz/column 6/.style={column sep=1em},
              /tikz/column 8/.style={column sep=-1.8em},
              /tikz/column 10/.style={column sep=1em},
              },
legend columns=6,
]

\pgfdeclarepattern{
  name=hatch,
  parameters={\hatchsize,\hatchangle,\hatchlinewidth},
  bottom left={\pgfpoint{-.1pt}{-.1pt}},
  top right={\pgfpoint{\hatchsize+.1pt}{\hatchsize+.1pt}},
  tile size={\pgfpoint{\hatchsize}{\hatchsize}},
  tile transformation={\pgftransformrotate{\hatchangle}},
  code={
    \pgfsetlinewidth{\hatchlinewidth}
    \pgfpathmoveto{\pgfpoint{-.1pt}{-.1pt}}
    \pgfpathlineto{\pgfpoint{\hatchsize+.1pt}{\hatchsize+.1pt}}
    \pgfpathmoveto{\pgfpoint{-.1pt}{\hatchsize+.1pt}}
    \pgfpathlineto{\pgfpoint{\hatchsize+.1pt}{-.1pt}}
    \pgfusepath{stroke}
  }
}

\tikzset{
  hatch size/.store in=\hatchsize,
  hatch angle/.store in=\hatchangle,
  hatch line width/.store in=\hatchlinewidth,
  hatch size=5pt,
  hatch angle=0pt,
  hatch line width=.5pt,
}

%\addplot[ybar,ybar legend,draw=black,fill=limegreen3122331,line width=0.08pt]
%  table[row sep=crcr]{%
%	0	0\\
%};
%\addlegendentry{Fixed Policy}

\addplot[ybar,ybar legend,draw=black,fill=gainsboro202218233,line width=0.08pt]
table[row sep=crcr]{%
	0	0\\
};
\addlegendentry{$C_{\rm HAP} = 3\, 000$ GFLOPS}

\addplot[ybar,ybar legend,draw=black,fill=cornflowerblue119170200,line width=0.08pt]
table[row sep=crcr]{%
	0	0\\
};
\addlegendentry{$C_{\rm HAP} = 4\, 000$ GFLOPS}

\addplot[ybar,ybar legend,draw=black,fill=steelblue50110161,line width=0.08pt]
  table[row sep=crcr]{%
	0	0\\
};
\addlegendentry{$C_{\rm HAP} = 5\, 000$ GFLOPS}

\addplot[ybar,ybar legend,draw=black,fill=black,line width=0.08pt,postaction={pattern=north east lines}, pattern color=white, yshift = 0.05cm]
table[]{%
	0	0
};
\addlegendentry{}
\addplot[color=black, dashdotted, very thick, yshift = -0.15cm]
table[]{%
	0	0
};
\addlegendentry{Baseline}

\addplot[ybar,ybar legend,draw=black,fill=chocolate2238331,line width=0.08pt,postaction={
	pattern=dots}]
table[row sep=crcr]{%
	0	0\\
};
\addlegendentry{Fully local}

\end{axis}
\end{tikzpicture}%
	\end{subfigure}
	%\vskip 0.5cm
	\centering
	\subfloat[][Real-time probability at the optimal offloading factor $\eta^*$.]
	{
		\label{fig:success}
		% This file was created with tikzplotlib v0.10.1.
\begin{tikzpicture}[scale=1]

%\definecolor{cornflowerblue119170200}{RGB}{119,170,200}
%\definecolor{darkslategray38}{RGB}{38,38,38}
%\definecolor{darkslategray66}{RGB}{66,66,66}
%\definecolor{gainsboro202218233}{RGB}{202,218,233}
%\definecolor{lightgray204}{RGB}{204,204,204}
%\definecolor{steelblue50110161}{RGB}{50,110,161}

\definecolor{lightgray204}{RGB}{204,204,204}
\definecolor{darkslategray38}{RGB}{38,38,38}
\definecolor{darkslategray66}{RGB}{66,66,66}
\definecolor{gainsboro202218233}{RGB}{202,218,233}
\definecolor{cornflowerblue119170200}{RGB}{119,170,200}
\definecolor{steelblue50110161}{RGB}{50,110,161}
\definecolor{red}{RGB}{223,83,31}

\begin{axis}[
width = \textwidth/2.1,
height = 4.5cm,
axis line style={lightgray204},
tick align=outside,
unbounded coords=jump,
x grid style={lightgray204},
xlabel=\textcolor{darkslategray38}{Frame rate $r$ [fps]},
xmajorticks=true,
xmin=-0.5, xmax=3.5,
xtick style={color=darkslategray38},
xtick={0,1,2,3},
xticklabels={5,10,15,20},
y grid style={lightgray204},
ylabel=\textcolor{darkslategray38}{Real-time probability $P_{\rm RT}(\eta^*)$ [\%]},
ymajorgrids,
ymajorticks=true,
ymin=0, ymax=104.999999993056,
ytick style={color=darkslategray38}
]

% \pgfdeclarepattern{
%   name=hatch,
%   parameters={\hatchsize,\hatchangle,\hatchlinewidth},
%   bottom left={\pgfpoint{-.1pt}{-.1pt}},
%   top right={\pgfpoint{\hatchsize+.1pt}{\hatchsize+.1pt}},
%   tile size={\pgfpoint{\hatchsize}{\hatchsize}},
%   tile transformation={\pgftransformrotate{\hatchangle}},
%   code={
%     \pgfsetlinewidth{\hatchlinewidth}
%     \pgfpathmoveto{\pgfpoint{-.1pt}{-.1pt}}
%     \pgfpathlineto{\pgfpoint{\hatchsize+.1pt}{\hatchsize+.1pt}}
%     \pgfpathmoveto{\pgfpoint{-.1pt}{\hatchsize+.1pt}}
%     \pgfpathlineto{\pgfpoint{\hatchsize+.1pt}{-.1pt}}
%     \pgfusepath{stroke}
%   }
% }

\tikzset{
  hatch size/.store in=\hatchsize,
  hatch angle/.store in=\hatchangle,
  hatch line width/.store in=\hatchlinewidth,
  hatch size=5pt,
  hatch angle=0pt,
  hatch line width=.5pt,
}

%\addlegendimage{ybar,ybar legend,draw=black,fill=gainsboro202218233,line width=0.08pt}
%\addlegendentry{3000GFLOPs}
\draw[draw=black,fill=gainsboro202218233,line width=0.08pt] (axis cs:-0.375,0) rectangle (axis cs:-0.225,99.9999998747832);
\draw[draw=black,fill=gainsboro202218233,line width=0.08pt] (axis cs:0.625,0) rectangle (axis cs:0.775,94.6240393662854);
\draw[draw=black,fill=gainsboro202218233,line width=0.08pt] (axis cs:1.625,0) rectangle (axis cs:1.775,39.646280942042);
\draw[draw=black,fill=gainsboro202218233,line width=0.08pt] (axis cs:2.625,0) rectangle (axis cs:2.775,0);

%\addlegendimage{ybar,ybar legend,draw=black,fill=cornflowerblue119170200,line width=0.08pt}
%\addlegendentry{4000GFLOPs}

\draw[draw=black,fill=cornflowerblue119170200,line width=0.08pt] (axis cs:-0.175,0) rectangle (axis cs:-0.025,99.9999999933826);
\draw[draw=black,fill=cornflowerblue119170200,line width=0.08pt] (axis cs:0.825,0) rectangle (axis cs:0.975,99.5236470124666);
\draw[draw=black,fill=cornflowerblue119170200,line width=0.08pt] (axis cs:1.825,0) rectangle (axis cs:1.975,58.2905459500849);
\draw[draw=black,fill=cornflowerblue119170200,line width=0.08pt] (axis cs:2.825,0) rectangle (axis cs:2.975,0);

%\addlegendimage{ybar,ybar legend,draw=black,fill=steelblue50110161,line width=0.08pt}
%\addlegendentry{5000GFLOPs}

\draw[draw=black,fill=steelblue50110161,line width=0.08pt] (axis cs:0.025,0) rectangle (axis cs:0.175,99.9999999726524);
\draw[draw=black,fill=steelblue50110161,line width=0.08pt] (axis cs:1.025,0) rectangle (axis cs:1.175,99.9999305684764);
\draw[draw=black,fill=steelblue50110161,line width=0.08pt] (axis cs:2.025,0) rectangle (axis cs:2.175,75.2484056757748);
\draw[draw=black,fill=steelblue50110161,line width=0.08pt] (axis cs:3.025,0) rectangle (axis cs:3.175,41.6145581477572);

%\addlegendimage{ybar,ybar legend,draw=black,fill=crimson2233131,line width=0.08pt}
%\addlegendentry{Fully-local}

\draw[draw=black,fill=red,line width=0.08pt,postaction={
	pattern=dots}] (axis cs:0.225,0) rectangle (axis cs:0.375,95.7864033966431);
\draw[draw=black,fill=red,line width=0.08pt,postaction={
	pattern=dots}] (axis cs:1.225,0) rectangle (axis cs:1.375,34.3083334788147);
\draw[draw=black,fill=red,line width=0.08pt,postaction={
	pattern=dots}] (axis cs:2.225,0) rectangle (axis cs:2.375,0);
\draw[draw=black,fill=red,line width=0.08pt,postaction={
	pattern=dots}] (axis cs:3.225,0) rectangle (axis cs:3.375,0);

% Baseline bars

\draw[draw=black,fill=black,line width=0.08pt,postaction={pattern=north east lines}, pattern color=gainsboro202218233] (axis cs:-0.325,0) rectangle (axis cs:-0.275,99.1217347235927);
\draw[draw=black,fill=black,line width=0.08pt,postaction={pattern=north east lines}, pattern color=cornflowerblue119170200] (axis cs:-0.125,0) rectangle (axis cs:-0.075,99.4060119382777);
\draw[draw=black,fill=black,line width=0.08pt,postaction={pattern=north east lines}, pattern color=steelblue50110161] (axis cs:0.075,0) rectangle (axis cs:0.125,99.5622303489139);

\draw[draw=black,fill=black,line width=0.08pt,postaction={pattern=north east lines}, pattern color=gainsboro202218233] (axis cs:0.675,0) rectangle (axis cs:0.725,76.1143175790264);
\draw[draw=black,fill=black,line width=0.08pt,postaction={pattern=north east lines}, pattern color=cornflowerblue119170200] (axis cs:0.875,0) rectangle (axis cs:0.925,81.4347576746435);
\draw[draw=black,fill=black,line width=0.08pt,postaction={pattern=north east lines}, pattern color=steelblue50110161] (axis cs:1.075,0) rectangle (axis cs:1.125,84.7632716427251);

\draw[draw=black,fill=black,line width=0.08pt,postaction={pattern=north east lines}, pattern color=gainsboro202218233] (axis cs:1.675,0) rectangle (axis cs:1.725,35.7602261627391);
\draw[draw=black,fill=black,line width=0.08pt,postaction={pattern=north east lines}, pattern color=cornflowerblue119170200] (axis cs:1.875,0) rectangle (axis cs:1.925,42.9999625809896);
\draw[draw=black,fill=black,line width=0.08pt,postaction={pattern=north east lines}, pattern color=steelblue50110161] (axis cs:2.075,0) rectangle (axis cs:2.125,47.9999999943214);

\draw[draw=black,fill=black,line width=0.08pt,postaction={pattern=north east lines}, pattern color=gainsboro202218233] (axis cs:2.675,0) rectangle (axis cs:2.725,0);
\draw[draw=black,fill=black,line width=0.08pt,postaction={pattern=north east lines}, pattern color=cornflowerblue119170200] (axis cs:2.875,0) rectangle (axis cs:2.925,0);
\draw[draw=black,fill=black,line width=0.08pt,postaction={pattern=north east lines}, pattern color=steelblue50110161] (axis cs:3.075,0) rectangle (axis cs:3.125,41.5532965744178);

\addplot [line width=1.08pt, darkslategray66]
table {%
	-0.3 nan
	-0.3 nan
};
\addplot [line width=1.08pt, darkslategray66]
table {%
	0.7 nan
	0.7 nan
};
\addplot [line width=1.08pt, darkslategray66]
table {%
	1.7 nan
	1.7 nan
};
\addplot [line width=1.08pt, darkslategray66]
table {%
	2.7 nan
	2.7 nan
};
\addplot [line width=1.08pt, darkslategray66]
table {%
	-0.1 nan
	-0.1 nan
};
\addplot [line width=1.08pt, darkslategray66]
table {%
	0.9 nan
	0.9 nan
};
\addplot [line width=1.08pt, darkslategray66]
table {%
	1.9 nan
	1.9 nan
};
\addplot [line width=1.08pt, darkslategray66]
table {%
	2.9 nan
	2.9 nan
};
\addplot [line width=1.08pt, darkslategray66]
table {%
	0.1 nan
	0.1 nan
};
\addplot [line width=1.08pt, darkslategray66]
table {%
	1.1 nan
	1.1 nan
};
\addplot [line width=1.08pt, darkslategray66]
table {%
	2.1 nan
	2.1 nan
};
\addplot [line width=1.08pt, darkslategray66]
table {%
	3.1 nan
	3.1 nan
};
\addplot [line width=1.08pt, darkslategray66]
table {%
	0.3 nan
	0.3 nan
};
\addplot [line width=1.08pt, darkslategray66]
table {%
	1.3 nan
	1.3 nan
};
\addplot [line width=1.08pt, darkslategray66]
table {%
	2.3 nan
	2.3 nan
};
\addplot [line width=1.08pt, darkslategray66]
table {%
	3.3 nan
	3.3 nan
};
\end{axis}

\end{tikzpicture}
	}
	\subfloat[][Statistics of the average latency for $r=10$ fps.]
	{
		\label{fig:cdf}
		\input{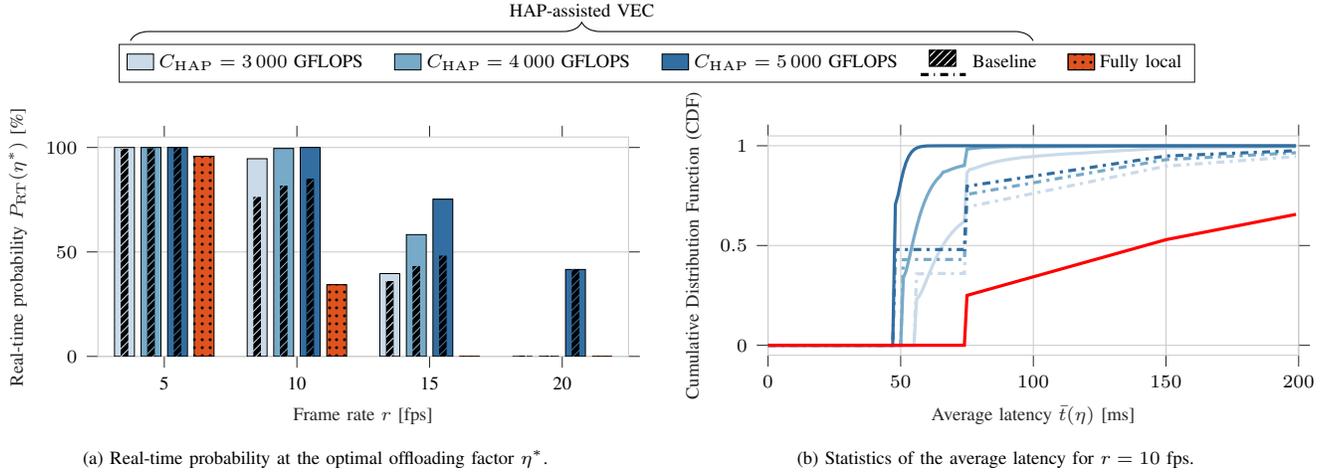}
	}
	\caption{Real-time probability (left) and average latency (right) vs. $C_{\rm HAP}$ and $r$, when $C_{\rm GV}=$ 800~GFLOPS, $n=100$~GVs and $n_\text{UL}=1$~Mb. Dash-dot lines and black striped bars refer to a baseline offloading scheme which balances the load of the HAP and of the GVs equally.\vspace{-0.5cm}}
	\label{fig:analysis_rate}
\end{figure*}
	
We analyze a rural/remote scenario with $n\in \{1, ..., 200\}$ \glspl{gv}, uniformly distributed over a large \gls{aoi} of 1000~$\text{km}^2$. 
The coverage area of the HAP is large enough to ensure that \glspl{gv} are always and continuously under coverage in the \gls{aoi}, even in case of mobility
Each \gls{gv} generates video frames of size $n_\text{UL} \in [1,3]$~Mb from its camera sensor at an average rate $r=10$~fps. Object detection on these frames requires a constant computational load of $C = 60$~GFLOP per frame, which is computed as the average between the computational performance of two popular object detectors, namely Gaussian YOLO and SqueezeDet+~\cite{li2021confidence}.
If frames are offloaded to the HAP (with probability $\eta^*$), eventually the processed output (i.e., the bounding boxes of the detected objects) is returned to the GVs in a packet of a much smaller size than the original frame, i.e., $n_\text{DL} = 100$~kb, which implies that $t_{\rm DL}\ll t_{\rm UL}$.
	
In the simulations, we compare a fully-local scheme in which all data frames are processed onboard (i.e., $\eta=0$) and the optimal offloading policy (where $\eta=\eta^*$) based on the solution of Eq.~\eqref{eq:opt_problem} with $t_\text{max}=1/r$. 
\Glspl{gv} have a computational capacity of $C_{\rm GV}\in\{200, 600, 800, 1000\}$~GFLOPS, whereas the \gls{hap} operates via $c=15$ parallel servers, each of which offers a computational capacity of $C_{\rm HAP}\in \{3000, 4000, 5000\}$~GFLOPS, so as to simulate different computing~conditions. These values are consistent with the capacity of off-the-shelf computing units: for example, the Nvidia GeForce RTX 3080 Mobile CPU (GeForce GTX 1080 GPU), which is compatible with the space/power constraints onboard GVs (HAP), offers a capacity of 300 (9000) GFLOPS.
	
All devices operate at a carrier frequency of $f_c =38$~GHz (mmWave) and with a bandwidth $B=400$~MHz, while for a complete description of the channel parameters in Sec.~\ref{sub:channel} we refer the interested reader to~\cite[Table 1]{traspadini2022uavhapassisted}.

\subsection{Numerical Results}
	\label{sub:numerical_results}
	
	\paragraph{Number of users (GVs)}
	In~\cref{fig:density}, we evaluate the impact of the number of GVs in terms of VEC performance. We can see that, on average, fully local processing onboard the \glspl{gv}, with $C_{\rm GV}=800$ GFLOPS, requires about $200$ ms for each video frame, regardless of the value of $n$, which is not compatible with real-time services. 
	As expected, the average latency for processing data via the HAP grows with $n$, due to the more frequent offloading requests and the resulting populated queues as the number of GVs increases.
	In any case, HAP-assisted VEC with powerful processing can reduce the average latency compared to the fully local scenario (up to almost 5 times when $n=50$ GVs), despite the communication delay for uploading data frames to the HAP servers and for delivering the processed output to the end nodes.

	Interestingly, the optimal offloading probability $\eta^{*}$ decreases with the number of GVs.
	When $n < 150$, the best choice is to offload data with probability $\eta^*>0.7$. This approach allows to achieve, on average, real-time data processing at the frame rate of the sensors,  i.e., $\bar{t}(\eta)<t_{\rm max}$.
	%This approach further promotes energy efficiency: while \glspl{hap} are solar powered and can harvest the energy they use for both hovering and computation, \glspl{gv} are battery powered, for which data offloading may increase the vehicle's autonomy~\cite{v2x_uav}. 
    On the other hand, when $n \ge 150$, the average latency $\bar{t}(\eta)$ to/from the HAP alone exceeds the latency constraint, i.e., $P(t_\text{HAP}\leq t_\text{max})=0$.
    This is due to the fact that the more populated system may overload the available channel bandwidth, thus resulting in longer transmission delays, which makes fully local processing an increasingly desirable option: in these conditions, the optimal offloading factor is as low as $\eta^*=0.25$. %which  because, in this way, the load at the \glspl{gv} is reduced, thus maximizing the average real-time probability.
%	 and deployed with \glspl{gv} supplied with a computational capability $C_{\rm GV}$ equal to 900~GFLOPs and the perception rate $r$ is set to 10~fps.

%	The choice of the Fixed Policy, with these conditions, does not guarantee real-time service, even though the latency is close to the maximum constraint.
%	When 200~\glspl{gv} are in the system, the choice of the Fixed Policy of offloading 3~fps results in an instability of the \gls{hap} queue, the optimal choice is to offload only 2~fps, however real-time service cannot be met with these parameters.
	
\paragraph{GV's computational capacity}
%This parameter is important to understand the real benefits of \gls{vec} in a rural scenario.
We let $C_{\rm GV}$ vary from 200 to 1000 GFLOPS,
%\footnote{We reasonably expect that $C_{\rm GV}\leq 1000$ GFLOPS, due to the high price and energy consumption of processing units. Even ignoring the price-performance ratio, more powerful hardware would reduce the vehicle's travel mileage and autonomy~\cite{v2x_uav}.} 
as considered in~\cite{traspadini2022uavhapassisted}, in a scenario with $n=90$~GVs, and plot the optimal offloading factor $\eta^*$ in~\cref{fig:capacity}. We see that $\eta^*$ decreases as $C_{\rm GV}$ increases, specifically from 0.81 (200~GFLOPS) to 0.64 (1000~GFLOPS). Fully local processing becomes increasingly more attractive as vehicles incorporate more powerful processing hardware, to avoid additional delays for data offloading. Still, for the values of capacity in~\cref{fig:capacity}, HAP-assisted VEC remains a more convenient choice to achieve real-time performance, even though the gap in terms of average latency compared to the fully local scenario is only 17\% when $C_{\rm GV}=1000$ GFLOPS. Notably, \cref{fig:capacity} shows that fully local processing leads to queue instability if $C_{\rm GV}\leq600$ GFLOPS, which further motivates the need for offloading to HAP.
		
\paragraph{Frame rate}
In~\cref{fig:success}, we study the real-time probability at the optimal offloading factor $\eta^*$, as a function of the frame rate $r$ and the computational capacity of the HAP. In general, increasing $r$ allows the sensor to capture data at better resolution, which leads to more accurate autonomous driving perception, at the cost of a higher data rate and complexity.
As expected, $P_{\rm RT}(\eta^*)$ is a decreasing function of~$r$: for $C_{\rm HAP}=5000$ GFLOPS, $P_{\rm RT}$ ranges from $1$ at $r=10$ fps to $0.41$ at $r=20$ fps. For $C_{\rm HAP}<5000$ GFLOPS, the system is constrained by the capacity of the queues at the HAP servers; in particular, the system becomes unstable as $r\geq 20$ fps. Moreover, for the fully local configurations, $P_{\rm RT}>0.95$ only with $r=5$ fps and $C_{\rm GV}=800$ GFLOPS, an indication that the limited computational capacity at the \glspl{gv} is not enough to support automotive tasks at the frame rate of the sensors.

For comparison, we considered a simple baseline offloading scheme that balances the load of the HAP and the GVs using an offloading factor $\eta_{\rm bl}$ given by
\begin{equation}
\eta_{\rm bl} = \left(\frac{n \cdot C_{\rm GV}}{c \cdot C_{\rm HAP}}+1\right)^{-1},
\end{equation}
which guarantees $G_{\rm HAP}(\eta_{\rm bl}) / c = G_{\rm GV}(\eta_{\rm bl})$. 
Results in~\cref{fig:success} show that this baseline can support real-time processing with a probability higher than 0.85 as long as the frame rate is less than 10~fps, otherwise the system becomes unstable. On the other hand, our proposed framework is able to regulate the offloading factor based on the actual capacity of the network, and can offer real-time performance even in more congested scenarios.

In~\cref{fig:cdf}, we plot the \gls{cdf} of the average latency for $r=10$~fps.
HAP-assisted VEC, even with $C_{\rm HAP}=3\,000$~GFLOPs, achieves real-time processing (i.e., $\bar{t}(\eta)\leq1/r=100$ ms) with a probability higher than $0.94$, vs. $0.34$ for fully local processing.
The \gls{cdf} of the baseline shows a trend similar to the fully local scheme, but with a constant gain from $0.44$ to $0.54$ which depends on $C_{\rm HAP}$.
From~\cref{fig:cdf}, it is clear that the \gls{cdf} shows two steps: the first step (at $\bar{t}(\eta)\simeq 50$ ms) corresponds to the minimum processing time for HAP-assisted VEC, i.e., the time for downlink and uplink transmissions, and the processing time at the HAP; the second step (at $\bar{t}(\eta)\simeq 75$ ms) corresponds to the minimum processing time for fully-local processing, i.e., $C/C_{GV}\simeq 75$ ms.

\section{Conclusions and Future Works}
	\label{sec:conclusions_and_future_works}
In this paper, we analyzed a rural scenario where autonomous \glspl{gv} capture sensory data to perform real-time object detection, and addressed the following research question: \emph{``Is it feasible to offload and process self-driving-related tasks to HAPs?"}
For this purpose, we described both the \glspl{gv} and the \gls{hap} as queues that receive sensory data, and determined a closed-form expression for the average waiting time of data in those queues.
We then formalized a VEC optimization problem and compared the case in which data processing is performed onboard the GVs (fully local scenario) vs. the case in which a fraction of the processing load is offloaded to HAP servers. Simulation results showed that the limited computational capacity of GVs is generally not compatible with real-time operations, while an optimal offloading factor exists to minimize the processing time. In particular, real-time performance requires a computational capacity at the HAP higher than 3000 GFLOPS for a frame rate $r\leq10$ fps. 

In future work, we will extend our offloading optimization problem by incorporating the impact of power consumption.

	\section*{Appendix: Proof of Lemma 1}
	
Let us consider an M/D/$c$ queue with arrival rate $\lambda$ and service rate $\mu$. The state probability distribution fulfills Eq.~	\eqref{eq:p_j}. The average length of the queue $E[L_q]$ can be written~as
\begin{align}
E[L_q] = \sum_{k=c}^{M-1} p_k (k-c) + \sum_{k=M}^{+\infty} p_k (k-c). \label{eq:2}
\end{align}
Using the approximation in Eq.~\eqref{eq:p_j_approx}, the second summation in Eq.~\eqref{eq:2} can be rewritten as
\medmuskip=4mu
\thinmuskip=4mu
\thickmuskip=4mu
\begin{align}
\sum_{k=M}^{+\infty} p_k (k-c) &= \sum_{k=M}^{+\infty} p_M \tau^{-(k-M)} (k-c)\\
&= p_M \left[\sum_{j=0}^{+\infty} \tau^{-j} j + \sum_{j=0}^{+\infty} \tau^{-j} (M-c)\right]\\
\label{eq:21}
&= p_M \left[\sum_{j=0}^{+\infty} \tau^{-j} j + \frac{M-c}{1-\frac{1}{\tau}} \right].
\end{align}
\medmuskip=6mu
\thinmuskip=6mu
\thickmuskip=6mu
%where we need the to use the following result:
% The summation in Eq.~\eqref{eq:21} is equal to
% \begin{align}
%  \sum_{j=0}^{+\infty} \tau^{-j} j&=\sum_{k=1}^{+\infty}\sum_{j=k}^{+\infty} \tau^{-j} \\ 
%  &=  \sum_{k=1}^{+\infty}\left[\sum_{j=0}^{+\infty} \tau^{-j} - \sum_{j=0}^{k-1} \tau^{-j}\right]\\
%  \label{eq:10}
% &=\sum_{k=1}^{+\infty}\left[\frac{\left(\frac{1}{\tau}\right)^k}{1-\frac{1}{\tau}}\right]= \frac{\frac{1}{1-\frac{1}{\tau}}-1}{1-\frac{1}{\tau}}.
% \end{align}
The summation in Eq.~\eqref{eq:21} is equal to
\begin{align}
 \sum_{j=0}^{+\infty} \tau^{-j} j &= \sum_{k=1}^{+\infty}\sum_{j=k}^{+\infty} \tau^{-j} = \sum_{k=1}^{+\infty}\tau^{-k} \sum_{j=0}^{+\infty} \tau^{-j} \\
%&=\sum_{k=1}^{+\infty}\sum_{j=k}^{+\infty} \tau^{-j} \\ 
 &=  \sum_{k=1}^{+\infty}\left[\sum_{j=0}^{+\infty} \tau^{-j} - \sum_{j=0}^{k-1} \tau^{-j}\right]\\
 \label{eq:10}
&=\sum_{k=1}^{+\infty}\left[\frac{\left(\frac{1}{\tau}\right)^k}{1-\frac{1}{\tau}}\right]= \frac{\frac{1}{1-\frac{1}{\tau}}-1}{1-\frac{1}{\tau}}.
\end{align}
Finally, $E[L_q]$ can be written from (\ref{eq:2}), (\ref{eq:21}), and (\ref{eq:10}) as
\medmuskip=2mu
\thinmuskip=2mu
\thickmuskip=2mu
\begin{align}
E[L_q] &= %\sum_{k=c}^{M-1} p_k (k-c) + p_M \left[\sum_{j=0}^{+\infty} \tau^{-j} j + \frac{M-c}{1-\frac{1}{\tau}} \right] \\
\sum_{k=c}^{M-1} p_k (k-c) + p_M \left[\frac{M-c+\frac{1}{1-\frac{1}{\tau}}-1}{1-\frac{1}{\tau}}\right].
\end{align}
\medmuskip=6mu
\thinmuskip=6mu
\thickmuskip=6mu
where $p_0$, $p_1$, ..., $p_M$ are computed by a linear system, as explained in Section \ref{sub:queuing_model}. 
The average waiting time in the queue, $E[W_q]$, is derived from Little's Law, i.e., $E[W_q] = E[L_q]/\lambda$, so that we obtain the expression in Eq.~\eqref{eq:Wq1}.

\bibliographystyle{IEEEtran}
\bibliography{IEEEabrv,bibliography.bib}

% Generated by IEEEtran.bst, version: 1.14 (2015/08/26)
\begin{thebibliography}{10}
\providecommand{\url}[1]{#1}
\csname url@samestyle\endcsname
\providecommand{\newblock}{\relax}
\providecommand{\bibinfo}[2]{#2}
\providecommand{\BIBentrySTDinterwordspacing}{\spaceskip=0pt\relax}
\providecommand{\BIBentryALTinterwordstretchfactor}{4}
\providecommand{\BIBentryALTinterwordspacing}{\spaceskip=\fontdimen2\font plus
\BIBentryALTinterwordstretchfactor\fontdimen3\font minus
  \fontdimen4\font\relax}
\providecommand{\BIBforeignlanguage}[2]{{%
\expandafter\ifx\csname l@#1\endcsname\relax
\typeout{** WARNING: IEEEtran.bst: No hyphenation pattern has been}%
\typeout{** loaded for the language `#1'. Using the pattern for}%
\typeout{** the default language instead.}%
\else
\language=\csname l@#1\endcsname
\fi
#2}}
\providecommand{\BIBdecl}{\relax}
\BIBdecl

\bibitem{giordani2020towards}
M.~Giordani \emph{et~al.}, ``{Toward 6G networks: Use cases and
  technologies},'' \emph{IEEE Commun. Mag.}, vol.~58, no.~3, pp. 55--61, March
  2020.

\bibitem{Chaoub20216g}
A.~Chaoub \emph{et~al.}, ``{6G for Bridging the Digital Divide: Wireless
  Connectivity to Remote Areas},'' \emph{IEEE Wireless Commun.}, pp. 160--168,
  July 2021.

\bibitem{giordani2021non}
M.~Giordani and M.~Zorzi, ``{Non-Terrestrial Networks in the 6G Era: Challenges
  and Opportunities},'' \emph{IEEE Network}, vol.~35, no.~2, pp. 244--251, Dec.
  2021.

\bibitem{nguyen20216g}
D.~C. Nguyen \emph{et~al.}, ``{6G Internet of Things: A Comprehensive
  Survey},'' \emph{IEEE Internet of Things Journal}, vol.~9, no.~1, pp.
  359--383, Jan. 2022.

\bibitem{liu2021vehicular}
L.~Liu \emph{et~al.}, ``{Vehicular edge computing and networking: A survey},''
  \emph{Mob. Netw. Appl.}, vol.~26, no.~3, pp. 1145--1168, Jun 2021.

\bibitem{v2x_offloading}
A.~Belogaev \emph{et~al.}, ``{Cost-Effective V2X Task Offloading in
  MEC-Assisted Intelligent Transportation Systems},'' \emph{IEEE Access},
  vol.~8, pp. 169\,010--169\,023, Sept. 2020.

\bibitem{velez20205g}
G.~Velez \emph{et~al.}, ``{5G beyond 3GPP Release 15 for connected automated
  mobility in cross-border contexts},'' \emph{Sensors}, vol.~20, no.~22, p.
  6622, Nov. 2020.

\bibitem{ke2021edge}
M.~Ke \emph{et~al.}, ``{An Edge Computing Paradigm for Massive IoT Connectivity
  over High-Altitude Platform Networks},'' \emph{IEEE Wireless Commun.},
  vol.~28, no.~5, pp. 102--109, Jun. 2021.

\bibitem{hap_offloading}
D.~S. Lakew \emph{et~al.}, ``{Intelligent Offloading and Resource Allocation in
  HAP-Assisted MEC Networks},'' in \emph{ICTC}, 2021.

\bibitem{wang2020potential}
D.~Wang \emph{et~al.}, ``{The Potential of Multi-Layered Hierarchical
  Non-Terrestrial Networks for 6G: A Comparative Analysis Among Networking
  Architectures},'' \emph{IEEE Veh. Technol. Mag.}, vol.~16, no.~3, pp.
  99--107, Sep. 2021.

\bibitem{traspadini2022uavhapassisted}
A.~Traspadini \emph{et~al.}, ``{UAV/HAP-Assisted Vehicular Edge Computing in
  6G: Where and What to Offload?}'' \emph{EuCNC-6G Summit}, 2022.

\bibitem{lundgren2014vehicle}
M.~Lundgren \emph{et~al.}, ``{Vehicle self-localization using off-the-shelf
  sensors and a detailed map},'' in \emph{IEEE Intelligent Vehicles Symposium},
  2014.

\bibitem{rossi2021role}
V.~Rossi \emph{et~al.}, ``{On the Role of Sensor Fusion for Object Detection in
  Future Vehicular Networks},'' in \emph{EuCNC-6G Summit}, 2021.

\bibitem{38821}
3GPP, “Solutions for NR to support Non-Terrestrial Networks (NTN),” TR
  38.821 (Release 16), 2020.

\bibitem{giordani2020satellite}
M.~Giordani and M.~Zorzi, ``{Satellite Communication at Millimeter Waves: a Key
  Enabler of the 6G Era},'' \emph{IEEE ICNC}, Feb 2020.

\bibitem{tijms2003Afirstcourse}
H.~Tijms, \emph{{A First Course in Stochastic Models}}.\hskip 1em plus 0.5em
  minus 0.4em\relax Wiley Ed., 2003.

\bibitem{gegenfurtner1992praxis}
K.~R. Gegenfurtner, ``{PRAXIS: Brent’s algorithm for function
  minimization},'' \emph{Behav. Res. Methods Instrum. Comput.}, vol.~24, no.~4,
  pp. 560--564, Dec. 1992.

\bibitem{li2021confidence}
W.~Li and K.~Liu, ``{Confidence-Aware Object Detection Based on MobileNetv2 for
  Autonomous Driving},'' \emph{Sensors}, vol.~21, no.~7, p. 2380, March 2021.

\end{thebibliography}

\end{document}